\documentclass[a4paper]{article}

\usepackage{INTERSPEECH2022}
\usepackage{cite}
\usepackage{subcaption}
\newsavebox{\bigimage}
\usepackage{float}
\usepackage{enumitem}
 
\usepackage{hyperref}
\hypersetup{
    hidelinks, 
    colorlinks=false, 
}
\usepackage{graphicx}
\usepackage{multirow}
\usepackage{comment}
\usepackage{makecell} 

\title{From KAN to GR-KAN: Advancing Speech Enhancement with KAN-Based Methodology\\[1em]}
\name{Haoyang Li$^{1}$, Yuchen Hu$^{1}$, Chen Chen$^{1}$, Sabato Marco Siniscalchi$^{2}$, Songting Liu$^{1}$, Eng Siong Chng$^{1}$}
\address{
  $^{1}$Nanyang Technological University, Singapore\\
  $^{2}$University of Palermo, Italy\\[1em]
\thanks{\hspace{-2mm}\fontsize{7.3pt}{7.3pt}\selectfont\textbf{Email: li0078ng@e.ntu.edu.sg}
}
}

\begin{document}

\maketitle
\begin{abstract}
Deep neural network (DNN)-based speech enhancement (SE) usually uses conventional activation functions, which lack the expressiveness to capture complex multiscale structures needed for high-fidelity SE. Group-Rational KAN (GR-KAN), a variant of Kolmogorov-Arnold Networks (KAN), retains KAN’s expressiveness while improving scalability on complex tasks. We adapt GR-KAN to existing DNN-based SE by replacing dense layers with GR-KAN layers in the time-frequency (T-F) domain MP-SENet and adapting GR-KAN's activations into the 1D CNN layers in the time-domain Demucs. Results on Voicebank-DEMAND show that GR-KAN requires up to 4× fewer parameters while improving PESQ by up to 0.1. In contrast, KAN, facing scalability issues, outperforms MLP on a small-scale signal modeling task but fails to improve MP-SENet. We demonstrate the first successful use of KAN-based methods for consistent improvement in both time- and SoTA TF-domain SE, establishing GR-KAN as a promising alternative for SE.
\end{abstract}
\noindent\textbf{Index Terms}: Speech Enhancement, Kolmogorov-Arnold Networks.

\vspace{-3mm}\section{Introduction}\vspace{-1mm}
\label{sec:intro}
Speech enhancement (SE) reduces noise and distortion to improve speech clarity, benefiting applications like hearing aids, telecommunications and voice recognition systems. Traditional SE solutions are based on digital signal processing solutions, such as Wiener filtering \cite{lim1978all}, spectral subtraction \cite{boll1979suppression} and minimum mean squared error estimation\cite{ephraim1984speech}. However, these approaches fail to track non-stationary noises and introduce annoying artifacts. Deep Neural Network (DNN)-based SE methods have proven their superiority in more recent years \cite{xu2014regression, lu13_interspeech, qian2025sav, zhang2025mamba}. Broadly speaking, we can classify DNN-based SE methods under two categories: (i) time domain methods \cite{pascual2017segan, defossez2020real, kim2021se, chen2021time, wang2024selm, li2025speech}, and 
(ii) time-frequency (TF) domain methods \cite{fu2019metricgan, cao2022cmgan, zadorozhnyy2022scp, dang2022dpt, yin2022tridentse, chen2025selective}. Time-domain methods aim to predict clean waveform directly from noisy counterparts, with Demucs \cite{defossez2020real} being a standard reference technique. Demucs combines a convolutional encoder-decoder with LSTM layers for effective sequential modeling. TF-domain methods predict a clean TF-domain representation and recover a time domain waveform from it. MP-SENet \cite{lu2023explicit} is the state-of-the-art (SoTA) in this category, which utilizes dilated DenseNet \cite{pandey2020densely} and Transformer \cite{vaswani2017attention} blocks to predict clean phase and magnitude spectrum, followed by waveform reconstruction through the Inverse Short-Time Fourier Transform.

DNN-based SE methods predominantly rely on standard activation functions, such as GELU \cite{hendrycks2016gaussian}, Swish \cite{ramachandran2017searching}, ReLU \cite{defossez2020real}, PReLU \cite{he2015delving}, and Leaky ReLU \cite{lu2023explicit}. While effective, these functions may limit a model’s ability to capture the intricate non-linear structures in speech, for instance,  harmonic patterns and phase variations, which are critical for high-quality enhancement. In particular, piecewise linear functions like ReLU, Leaky ReLU, and PReLU struggle with modeling smooth variations, e.g., sinusoidal components \cite{szandala2021review}. Although smoother activations like GELU and Swish alleviate this issue to some extent, these activation function have a higher computational costs and do not consistently outperform ReLU across different tasks \cite{szandala2021review, ramachandran2017searching}. These limitations, which we will further support through experiments, suggest that DNN with conventional activation functions may not be optimal for learning the complex representations necessary for SE.

Kolmogorov-Arnold Networks (KANs) \cite{liu2024kan} have recently emerged as an alternative to MLPs due to their enhanced expressiveness, continual learning capability, and interpretability. Unlike traditional MLP, KAN consists entirely of learnable univariate activation functions on the edges, each parameterized by a spline. This structural modification, grounded in the Kolmogorov-Arnold theorem, allows KAN to theoretically model complex non-linear patterns more effectively than MLPs that use conventional activation functions. However, despite these theoretical advantages, KAN sometimes fails to scale to complicated problems in practice \cite{yang2024kolmogorov} \cite{yu2024kan}. Its use of independent spline functions per edge causes rapid parameter growth, and its weight initialization disregards variance-preserving principles, leading to unstable training dynamics \cite{yang2024kolmogorov}. Hence, previous attempt on adapting KAN to SE had limited success \cite{mai5metricgan+}. In \cite{mai5metricgan+}, replacing linear layers with KAN layers in Metricgan+ \cite{fu2021metricgan+}'s  generator generally degraded the overall performance. 

To address the above-mentioned KAN's limitations, a new KAN variant referred to Group-Rational KANs (GR-KANs) was proposed in \cite{yang2024kolmogorov}. GR-KANs follow the foundational idea of KANs but modify the way functions are learned within the network, namely rational functions are used as activation functions. Furthermore, a group-theoretic structure is imposed on the activations to improve computational efficiency and avoid excessive parameter growth. GR-KANs also use a variance-preserving weight initialization strategy to improve training stability over KAN. In this work, we explore GR-KAN for SE. We first compare KAN and GR-KAN on a small-scale synthetic signal modeling task and find that both outperform conventional MLPs. To further assess scalability, we have integrated either KAN or GR-KAN layers into the time-frequency (TF) SoTA MP-SENet model. This was done by replacing the dense layers in the TF-Transformer blocks with KAN or GR-KAN layers. The experimental results show that (i) KAN layers do not improve MP-SENet quality despite the use of more trainable parameters, and (ii) GR-KAN layers outperform dense layers with various conventional and learnable activation functions, maintaining a comparable or smaller parameter count. In addition, when integrated into the 1D CNN layers in the time-domain Demucs model, GR-KANs also enable superior performance with four times fewer parameters than the standard configuration. Our findings show that while KAN struggles to adapt to SE, its variant, GR-KAN, overcomes these limitations and can be easily integrated into current DNN-based SE models to improve model performance while maintaining/reducing model size.

\section{Preliminaries}
\subsection{KAN: Kolmogorov-Arnold Network}
The Kolmogorov-Arnold theorem \cite{kolmogorov1957representation} asserts that any continuous function can be represented as a composition of univariate continuous functions of a finite number of variables. A KAN layer $L$ is thus a composition of learnable univariate functions, $\phi$(s), as shown in Eq. (\ref{eq:KAN_layer}):

\begin{equation}
\begin{aligned}
    L(\mathbf{x}) = \begin{bmatrix}
    \sum_{i=1}^{I} \phi_{i,1}(x_i) & \dots & \sum_{i=1}^{I} \phi_{i,J}(x_i)
    \end{bmatrix}
\end{aligned}
\label{eq:KAN_layer}
\vspace{-1mm}
\end{equation}

\noindent where $I$ and $J$ are the input and output dimensions. In practice, $\phi$ is approximated by Eq. (\ref{eq:KAN_layer_bspline}):
\begin{equation}
\begin{aligned}
\phi(x) &= w_1 b(x) + w_2 S(x) \quad S(x) = \sum_i n_iB_i(x)
\end{aligned}
\label{eq:KAN_layer_bspline}
\vspace{-1mm}
\end{equation}
\noindent where $w_1$ and $w_2$ are learnable parameters, $b$ is a basis, such as Swish. $S$ is a spline function, where $B_i$ is the B-spline basis function associated with each trainable control point $n_i$.

While the replacement of the single scalar weight on each edge of MLP with a KAN activation function $\phi$ brings about greater expressiveness \cite{liu2024kan}, this also makes the KAN layer significantly larger and more computationally expensive. KANs were reported to be 10 times slower than MLPs in \cite{liu2024kan}. Additionally, \cite{yang2024kolmogorov} noted that KAN violates the variance-preserving principle, impairing its trainability and convergence.

\subsection{GR-KAN: KAN variant based on rational functions}
GR-KAN \cite{yang2024kolmogorov} replaces B-spline with a rational function, due to its superior efficiency and expressiveness from a theoretical standpoint. Furthermore, \cite{yang2024kolmogorov} splits the $I$ input channels into $k$ groups and shares the parameters of the rational functions across all channels in the same group to reduce the number of parameters. A GR-KAN layer follows Eq. (\ref{eq:GRKAN}):

\begin{equation}
\begin{aligned}
    L(\mathbf{x}) &= 
    \begin{bmatrix}
    \sum_{i=1}^{I} w_{i,1} F_{\lfloor \frac{i}{I_k} \rfloor}(x_i) & \dots & \sum_{i=1}^{I} w_{i,J} F_{\lfloor \frac{i}{I_k} \rfloor}(x_i)
    \end{bmatrix}
\end{aligned}
\label{eq:GRKAN}
\vspace{-1mm}
\end{equation}

 \noindent where the activation $\phi_{i,j}$ in Eq. \ref{eq:KAN_layer} becomes $w_{i,j} \times F_{\lfloor \frac{i}{I_k} \rfloor}$, with $w_{i,j}$ being a scaler, and $F_{\lfloor \frac{i}{I_k} \rfloor}$  a rational function. $I_k = I/k$ is the number of channels in each group. 
 A GR-KAN layer $L$ is practically implemented as in Eq. (\ref{eq:GRKAN_implement}):

 \begin{equation}
\begin{aligned}
    L(x) = LIN(GR(x))
\end{aligned}
\label{eq:GRKAN_implement}
\vspace{-1mm}
\end{equation}
\noindent where $LIN$ is the matrix of $w$s in Eq. (\ref{eq:GRKAN}), and $GR$ is the vector of rational functions. 

\section{KAN and GR-KAN in SE}
\subsection{Analysis on small-scale signal modeling}
\label{subsec: preliminary exp}

We first evaluate KAN and GR-KAN solutions on a small-scale signal modeling task using a 5 second synthetic signal with sampling rate of 100. Results are compared against several MLP variants, with either conventional or learnable activation functions. The synthetic signal consists of dynamic, artificial syllables (150–250ms) with irregular pauses (20–100ms). The base frequency of each syllable fluctuates nonlinearly around 5 Hz, modulated by sine and cosine functions for smooth transitions. The amplitude of each syllable is shaped by an exponential decay envelope, randomly scaled between 0.5 and 1.5. Three formant frequencies (500 Hz, 1500 Hz, 3000 Hz) are added to each syllable with slight modulation ($\pm$40 Hz) and random phase shifts to simulate speech-like resonances. Lastly, gaussian noise (0.05 std) is added to the overall signal to introduce natural imperfections. The result is a signal with fluctuating pitch, amplitude and phase dynamics, and added resonant components, loosely capturing aspects of speech dynamics.

\subsection{Analysis on T-F domain SE}
\label{subsec: KAN in TF SE}
\begin{figure}[!t]
    \centering
    \includegraphics[scale=0.40]{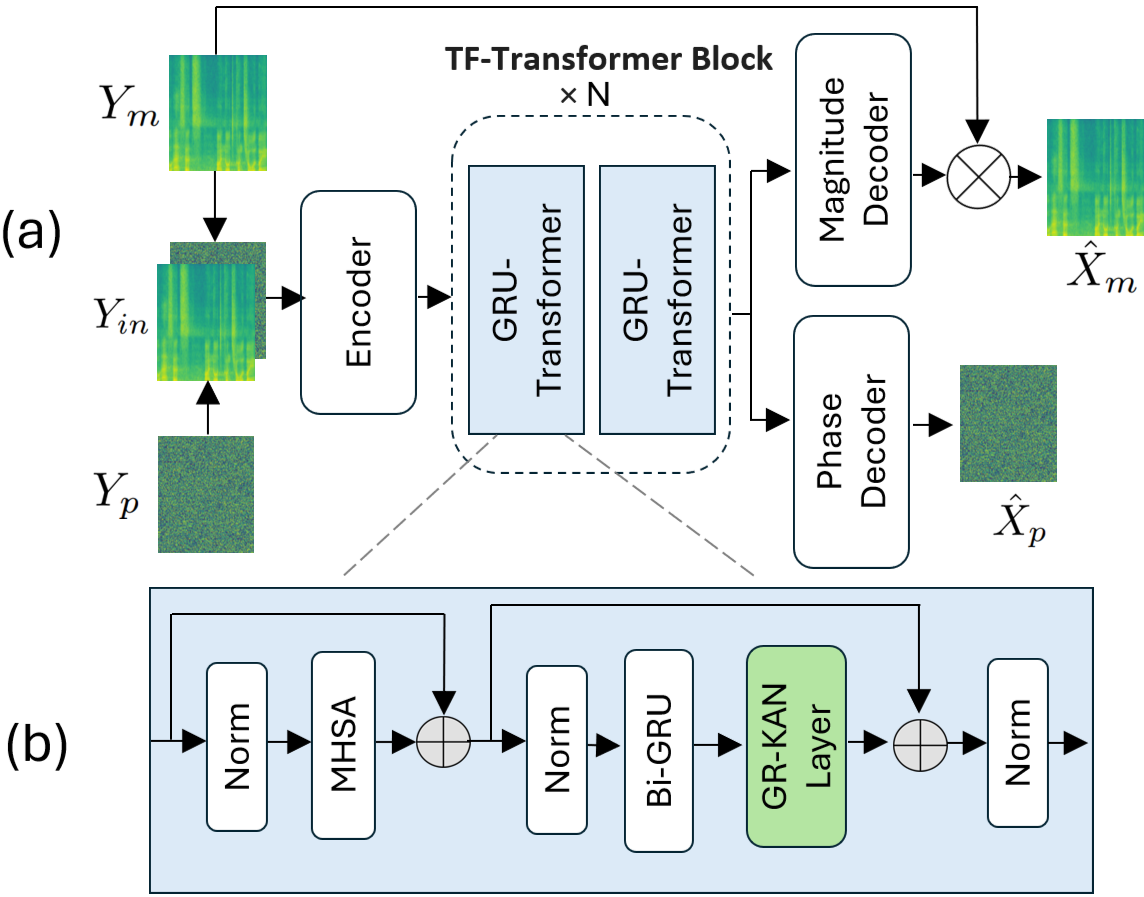}
    \caption{Architecture of (a) the Overall MP-SENet (b) the GR-KAN adapted GRU-Transformer Block.}
    \label{fig:mpsenet_kan} 
\end{figure}

 KAN and GR-KAN based SE solutions are compared using the TF-domain MP-SENet architecture. Figure \ref{fig:mpsenet_kan}a illustrates the overall architecture of the GR-KAN adapted model. The magnitude spectrum, $Y_m$, and the wrapped phase spectrum, $Y_p$, are stacked and fed into the MP-SENet encoder, followed by $N=4$ TF-Transformer blocks to capture local and global dependencies across the time and frequency dimensions. The output is then fed into a Magnitude Decoder and a Phase Decoder separately to restore the enhanced, magnitude spectrum $\hat X_m$ and the enhanced phase spectrum $\hat X_p$. We adapt GR-KAN to the GRU-Transformer blocks by adding a GR-KAN layer (same as Eq. \ref{eq:GRKAN_implement}) after the Bi-GRU block, as illustrated in figure \ref{fig:mpsenet_kan}b. To compare GR-KAN with KAN, we swap the GR-KAN layer with a KAN layer, where the KAN layer follows the implementation of efficient-kan \footnote{\url{https://github.com/Blealtan/efficient-kan}}. To further compare KAN and GR-KAN layer with conventional dense layers, we swap the GR-KAN layer in figure \ref{fig:mpsenet_kan}b with a linear layer preceded by an activation function such as GELU.

\subsection{Analysis on Time domain SE}
\begin{figure}[!t]
    \centering
    \includegraphics[scale=0.28]{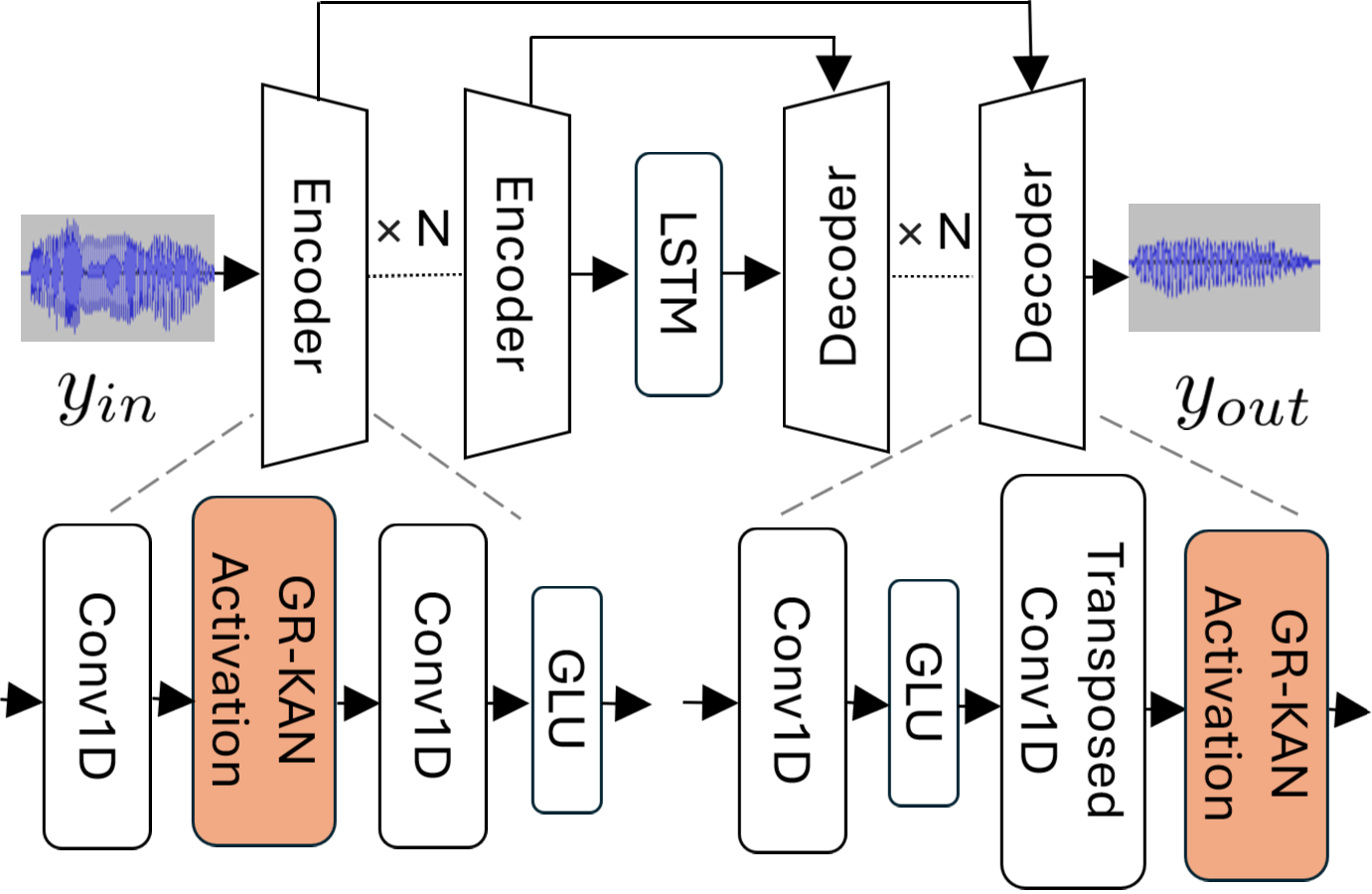}
    \caption{Architecture of the GR-KAN adapted Causal Demucs, where we replace all ReLU activations in the Encoder and Decoder blocks with GR-KAN activations. Please note that the last Decoder block does not have the GR-KAN activations.}
    \label{fig:demucs_kan} 
\end{figure}

To further assess the robustness of GR-KAN for general SE, we select the well-known time-domain SE model, Demucs, for adaptation, which operates in a different data domain than the TF-domain MP-SENet. Figure \ref{fig:demucs_kan} illustrates the GR-KAN adapted causal Demucs, where we have replaced the ReLU activation functions in the original Encoders and Decoders with the GR-KAN activation functions ($GR$ from Eq. \ref{eq:GRKAN_implement}). This setup differs slightly from GR-KAN's original formulation in Eq. (\ref{eq:GRKAN_implement}), since a 1D CNN layer is used instead of $LIN$ from Eq. (\ref{eq:GRKAN_implement}). To further access scalability, we compare our GR-KAN adapted Demucs with the original Demucs across different model sizes by varying the number of encoder and decoder blocks.

\section{Experiments}
\label{sec:Experiments}
\subsection{Experimental Setup}
\label{subsec: experiment setup}

The VoiceBank-DEMAND \cite{valentini2016investigating}, a widely recognized SE benchmark, is used to assess our KAN-based models. In this dataset, each clean utterance is paired with a corresponding noisy version. Following standard practice, all audio clips were downsampled to 16kHz. Finally, training and testing SNRs and noises do not match. More details can be found in \cite{valentini2016investigating}.

For the analysis indicated in Section \ref{subsec: preliminary exp}, 3 sequentially arranged linear layers with input and output dimensions of ($1$, $h$), ($h$, $h$) and ($h$, $1$), respectively, are used to compare GR-KAN and several MLP variants. An activation function is placed between every 2 linear layers. For GR-KAN, the GR-KAN activation ($GR$ from Eq. \ref{eq:GRKAN_implement}) is used as the activation function. For other MLP variants, ReLU, GELU, PAU \cite{molina2019pad}, and APL \cite{agostinelli2014learning} are used, where PAU and APL are learnable activation functions. For a relatively fair comparison in terms of model size, we set $h=8$ for GR-KAN and APL, and $h=12$ for other MLP based architectures. To further compare with KAN, 2 sequentially arranged KAN layers with input and output dimensions of ($1$, $4$) and ($4$, $1$), respectively, are used. The grid size and spline order of the KAN layers are set to 5 and 3, respectively. All models are trained for 300k epochs with learning rate of 0.001. The adam optimizer, and the mean squared error loss were used. 

For MP-SENet, we used a hop size, Hanning window size and FFT point number of 100, 400 and 400, respectively. All MLP and GR-KAN models were trained for 200 epochs with a batch size of 4. However, for all KAN models, the batch size was reduced to 2 to prevent GPU out-of-memory (OOM) errors. AdamW Optimizer \cite{loshchilov2017decoupled} with $\beta$1 = 0.8 and $\beta$2 = 0.99 was used. The learning rate was set to 0.0005, with a decaying factor of 0.99 every epoch. Our loss functions are identical to the original work \cite{lu2023explicit}. For GR-KAN, the group size $k$ is set to 8. For KAN, we set the grid size to 5 and the spline order to 3. For APL activation, we set the number of learnable negative slopes to 5 and introduced an L2 penalty to regularize $a^{s}_i$ and $b^{s}_i$ identically to the original work \cite{agostinelli2014learning}. 

The causal Demucs with depth  $n \in \{4, 5, 6\}$ (i.e. $n$ encoder blocks and $n$ decoder blocks) and a 2-layer unidirectional LSTM was used. We set the initial hidden dimension to 48, kernel size to 8, stride size and resample factor to 4. All models are trained for 500 epochs using the adam optimizer at a learning rate of 0.0003 and batch size of 16. The models are optimized using L1 loss on the waveform and a multi-resolution STFT loss on the magnitude spectrum.

\subsection{Evaluation Metrics}
Following \cite{fu2019metricgan}, we evaluated our models using PESQ \cite{rix2001perceptual}, CSIG, CBAK, COVL \cite{hu2007evaluation} and STOI \cite{taal2011algorithm}, which measure perceptual speech quality, signal distortion, noise distortion, overall quality and speech intelligibility, respectively. Higher scores indicate better performance. We also report the \#P, the number of parameters in the models.

\subsection{Experimental Results}
\begin{figure*}[!t]
    \centering
    \includegraphics[scale=0.52]{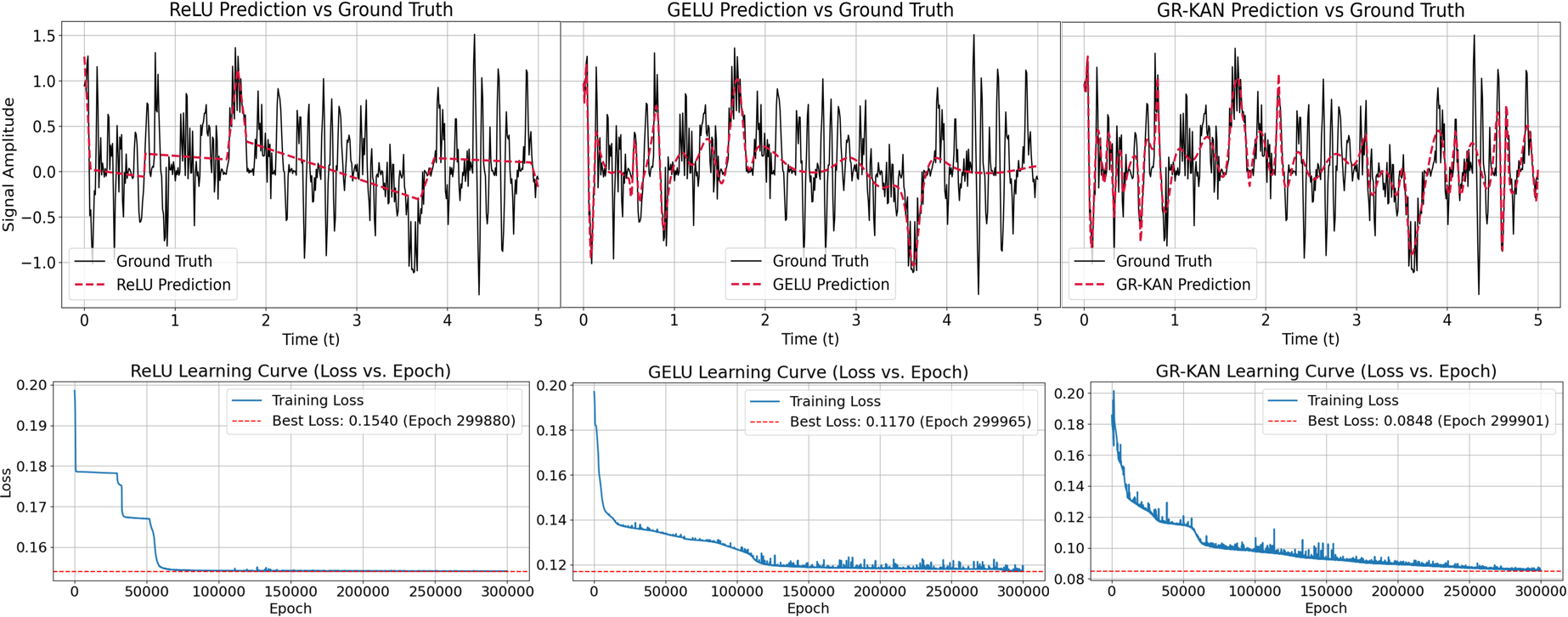}
    \caption{Comparison of MLP (ReLU), MLP (GELU) and GR-KAN on fitting an artificial signal with speech dynamics}
    \label{fig:basic_comparison} 
\end{figure*}

Table 1 presents the results, in terms of Mean Squared Error (MSE), of the signal fitting task described in Section \ref{subsec: preliminary exp}. In this small-scale task, both KAN and GR-KAN outperform the four MLP variants, whether using fixed activation functions (ReLU, and GELU) or learnable ones (PAU, and APL). This results seems to suggest that KAN-based models have higher expressiveness than MLPs. We also visualize the artificial signal with speech dynamics and some of the function fitting results in figure \ref{fig:basic_comparison}. From the figure, the ReLU-activated MLP struggle with smooth curvature modeling due to its piecewise linear nature. The GELU-activated MLP smooths out oscillations from 2s onward, losing fine-grained variations. In contrast, GR-KAN preserves more fine-grained variations and aligns peaks and valleys more accurately. These results suggest that GR-KAN holds strong potential for speech modeling tasks, where preserving subtle acoustic details and capturing dynamic, nonlinear patterns are critical.

\renewcommand{\arraystretch}{1.2}
\begin{table}[hbt!]
\caption{Comparison between KAN, GR-KAN and several MLP variants on the artificial signal modeling task}
\label{tab:speech_signal_modeling_result}
\resizebox{0.48\textwidth}{!}{
\begin{tabular}{c c c c c c c}\hline
\multirow{2}{*}{\begin{tabular}[c]{@{}c@{}} \end{tabular}} & 
\multirow{2}{*}{\begin{tabular}[c]{@{}c@{}}ReLU\end{tabular}} & 
\multirow{2}{*}{\begin{tabular}[c]{@{}c@{}}GELU\end{tabular}}  & 
\multirow{2}{*}{\begin{tabular}[c]{@{}c@{}}PAU\end{tabular}} & 
\multirow{2}{*}{\begin{tabular}[c]{@{}c@{}}APL\end{tabular}}  & 
\multirow{2}{*}{\begin{tabular}[c]{@{}c@{}}GR-KAN\end{tabular}}  & 
\multirow{2}{*}{\begin{tabular}[c]{@{}c@{}}KAN\end{tabular}} \\
 & & & & \\
\hline
MSE & 0.154 & 0.117 & 0.090 & 0.121 & 0.085 & \textbf{0.081} \\
\#P & 193 & 193 & 213 & 257 & \textbf{173} & 240 \\
\hline
\end{tabular}
}
\vspace{2mm}
\end{table}

Table \ref{tab:mpsenet_transformer_result} reports the performance of MP-SENet models using KAN layers, GR-KAN layers, or dense layers in the GRU-Transformer blocks. To account for variability from random initialization, all models were trained three times, and the final test performance was averaged. In this more complex task, KAN layers do not outperform dense layers with conventional activation functions, and they also increase the overall model size by approximately 50\%. This finding is consistent with previous studies \cite{yang2024kolmogorov} \cite{yu2024kan}, which suggest that KAN can struggle with more challenging tasks. GR-KAN layers instead consistently outperform dense layers with conventional and learnable activation functions with a similar parameter count of around 2.26M. Even when the number of dense layers is doubled (LeakyReLU*, GELU*, PReLU* in table \ref{tab:mpsenet_transformer_result}),  models using  conventional dense layers still consistently underperform the GR-KAN adapted model in PESQ and COVL despite having a higher parameter count. These findings demonstrate GR-KAN's stronger expressiveness and parameter efficiency against conventional MLP methods in speech enhancement.

\renewcommand{\arraystretch}{1.2}
\begin{table}[hbt!]
\caption{Comparison between KAN, GR-KAN and dense layers with  different activation functions in MP-SENet. * indicates that the number of dense layers used in the model are doubled. Performance reported in terms of mean and standard deviation over 3 runs. }
\label{tab:mpsenet_transformer_result}
\resizebox{0.47\textwidth}{!}{
\begin{tabular}{c c c c c}\hline
\multirow{2}{*}{\begin{tabular}[c]{@{}c@{}}Method\end{tabular}} & 
\multirow{2}{*}{\begin{tabular}[c]{@{}c@{}}PESQ\end{tabular}} & 
\multirow{2}{*}{\begin{tabular}[c]{@{}c@{}}COVL \end{tabular}}  & 
\multirow{2}{*}{\begin{tabular}[c]{@{}c@{}}STOI \end{tabular}}  & 
\multirow{2}{*}{\begin{tabular}[c]{@{}c@{}}\#P(M)\end{tabular}} \\
 & & & & \\
\hline
LeakyReLU & $3.557\pm0.002$ & $4.206\pm0.003$ & $0.958\pm0.000$ & 2.26 \\
GELU & $3.561\pm0.006$ & $4.202\pm0.006$ & $0.960\pm0.001$ & 2.26 \\
PReLU & $3.553\pm0.010$ & $4.204\pm0.014$ & $0.960\pm0.001$ & 2.27 \\
APL & $3.554\pm0.004$ & $4.203\pm0.009$ & $0.960\pm0.001$ & 2.28 \\
LeakyReLU* & $3.567\pm0.004$ & $4.218\pm0.003$ & $0.961\pm0.000$ & 2.30 \\
GELU* & $3.566\pm0.004$ & $4.212\pm0.006$ & $0.960\pm0.000$ & 2.30 \\
PReLU* & $3.573\pm0.009$ & $4.221\pm0.013$ & $0.960\pm0.001$ & 2.30 \\
KAN & $3.564\pm0.002$ & $4.213\pm0.011$ & $0.961\pm0.000$ & 3.44 \\
GR-KAN & $\textbf{3.588}\pm0.007$ & $\textbf{4.229}\pm0.007$ & $0.960\pm0.001$ & 2.26 \\
\hline
\end{tabular}
}
\vspace{2mm}
\end{table}

Table \ref{tab:demucs_result} reports Demucs's performance when the GR-KAN activation function is adapted to the encoder blocks only (KAN Enc), decoder blocks only (KAN Dec) or both. For all 3 adaptations, GR-KAN consistently improve the original Demucs in all metrics, with up to 0.1 increase in PESQ. This demonstrates that the incorporation of the GR-KAN activations enhances the model's ability to capture complex dependencies, resulting in improved performance, regardless of whether the adaptation is applied to the encoder, decoder, or both components.
\renewcommand{\arraystretch}{1.2}
\begin{table}[hbt!]
\centering
\caption{Results of the GR-KAN adapted Demucs at depth $5$.}
\label{tab:demucs_result}
\resizebox{0.48\textwidth}{!}{
\begin{tabular}{c c c c c c c}\hline
\multirow{2}{*}{\begin{tabular}[c]
{@{}c@{}}\makecell{\\ KAN \\ Enc}\end{tabular}} & 
\multirow{2}{*}{\begin{tabular}[c]
{@{}c@{}}\makecell{\\ KAN \\ Dec}\end{tabular}} & 
\multirow{2}{*}{\begin{tabular}[c]{@{}c@{}}\makecell{\\ PESQ}\end{tabular}} & 
\multirow{2}{*}{\begin{tabular}[c]{@{}c@{}}\makecell{\\ CSIG} \end{tabular}}  & 
\multirow{2}{*}{\begin{tabular}[c]{@{}c@{}}\makecell{\\ CBAK} \end{tabular}}  & 
\multirow{2}{*}{\begin{tabular}[c]{@{}c@{}}\makecell{\\ COVL} \end{tabular}}  & 
\multirow{2}{*}{\begin{tabular}[c]{@{}c@{}}\makecell{\\ STOI} \end{tabular}} \\
 & & & & \\
 \\
\hline
N & N & 2.896 & 4.284 & 3.429 & 3.608 & 0.945 \\
Y & N & 2.975 & 4.348 & 3.498 & 3.683 & 0.947 \\
N & Y & \textbf{2.990} & \textbf{4.349} & 3.495 & \textbf{3.695} & 0.947 \\
Y & Y & 2.987 & 4.342 & \textbf{3.500} & 3.688 & 0.947 \\
\hline
\end{tabular}
}
\end{table}

Table \ref{tab:demucs_scaling_result} compares the GR-KAN adapted Demucs and the original Demucs at different depth levels. The GR-KAN adapted Demucs consistently outperforms the original Demucs at the same depth level. In addition, the GR-KAN adapted Demucs at depth 5 outperforms the original model at depth 6, despite the latter having more than four times the total number of parameters. The results further demonstrate the superior expressiveness and parameter efficiency of the GR-KAN activation functions in the time-domain Demucs. 

\renewcommand{\arraystretch}{1.2}
\begin{table}[hbt!]
\caption{Comparison of the GR-KAN adapted Demucs and the original Demucs at different depth level}
\label{tab:demucs_scaling_result}
\resizebox{0.48\textwidth}{!}{
\begin{tabular}{c c c c c c c}\hline
\multirow{2}{*}{\begin{tabular}[c]{@{}c@{}}KAN\end{tabular}} & 
\multirow{2}{*}{\begin{tabular}[c]{@{}c@{}}Depth\end{tabular}}  & 
\multirow{2}{*}{\begin{tabular}[c]{@{}c@{}}PESQ\end{tabular}} & 
\multirow{2}{*}{\begin{tabular}[c]{@{}c@{}}CBAK \end{tabular}}  & 
\multirow{2}{*}{\begin{tabular}[c]{@{}c@{}}COVL \end{tabular}}  & 
\multirow{2}{*}{\begin{tabular}[c]{@{}c@{}}STOI \end{tabular}}  & 
\multirow{2}{*}{\begin{tabular}[c]{@{}c@{}}\#P(M)\end{tabular}} \\
 & & & & \\
\hline
N & 4 & 2.822 & 3.387 & 3.543 & 0.944
& 4.702 \\
Y & 4 & 2.885 & 3.426 & 3.596 & 0.944 & 4.702 \\
N & 5 & 2.896 & 3.429 & 3.608 & 0.945
& 18.868 \\
Y & 5 & 2.990 & 3.495 & 3.695 & 0.947
& 18.868 \\
N & 6 & 2.977 & 3.488 & 3.689 & 0.948
& 75.512 \\
Y & 6 & \textbf{3.018} & \textbf{3.511} & \textbf{3.721} & 0.948
& 75.512 \\
\hline
\end{tabular}
}
\vspace{2mm}
\end{table}

\section{Conclusion}
This work explores the use of KAN and its variant, GR-KAN, to enhance existing DNN-based SE solutions. We begin by demonstrating the superior expressiveness of KAN-based methods over MLPs with conventional and learnable activation functions through a small-scale signal modeling task. We then explain KAN's inability to scale to complex SE task, supported by experiments on MP-SENet. By integrating GR-KAN, a KAN variant designed to overcome these scalability challenges into MP-SENet and Demucs, we achieve consistent performance improvements across time-domain and time-frequency domain SE models, while requiring up to 4 times fewer trainable parameters. These promising results suggest that future SE methods and other speech generation models may benefit from adopting GR-KAN to enhance performance.

\newpage
\bibliographystyle{IEEEtran}
\bibliography{refs}

\end{document}